%% file: ICASSP_RIS.tex
\documentclass{article}
\usepackage{spconf,amsmath,graphicx}
\usepackage{amssymb}
\usepackage{algpseudocode}
\usepackage{cite}
\usepackage{amsfonts}
\usepackage{graphicx,subfigure}
\usepackage{fancyhdr}  %
\usepackage{cases}
\usepackage{extarrows}
\usepackage{algorithm}
\usepackage{multirow,tabularx}
\usepackage{mathtools}

\usepackage{color,xcolor}
\usepackage{etoolbox}
\usepackage{glossaries}

\input{abbreviations}

\input{acronyms}

\title{Energy Efficiency Maximization in RIS-Aided Networks\\ with Global Reflection Constraints}
\name{Robert K. Fotock$^{1,2}$, Alessio Zappone$^{1,2}$, Marco Di Renzo$^{3}$ 
\thanks{The authors have been supported by the European Commission through the H2020-MSCA-ITN-METAWIRELESS project, grant agreement 956256. The work of M. Di Renzo was also supported in part by the European Commission through the H2020 ARIADNE project, grant agreement 871464 and the H2020 RISE-6G project, grant agreement 101017011.}
}
\address{1: Consorzio Nazionale Interuniversitario per le Telecomunicazioni, Italy\\
2: University of Cassino and Southern Lazio, Italy \\
3: Universit\'e Paris-Saclay, CNRS, CentraleSup\'elec, Laboratoire des Signaux et Syst\`emes, France}
\begin{document}
%
\maketitle
\begin{abstract}
This work addresses the issue of energy efficiency maximization in a multi-user network aided by a reconfigurable intelligent surface (RIS) with global reflection capabilities. Two optimization methods are proposed to optimize the mobile users' powers, the RIS coefficients, and the linear receive filters. Both methods are provably convergent and require only the solution of convex optimization problems. The numerical results show that the proposed methods largely outperform heuristic resource allocation schemes. 
\end{abstract}
\vspace{-0.1cm}
\section{Introduction}\label{Sec:Intro}
\vspace{-0.1cm}
Reconfigurable intelligent surfaces (RISs) are emerging as a promising technology for future 6G wireless networks \cite{Ref10,RuiZhang_COMMAG,SmartWireless,HuangMag2020}. Besides providing a large number of degrees of freedom for signal transmission, RISs are particularly attractive from an energy-efficient point of view for their nearly passive behavior. Energy efficiency (EE) was already considered a key performance indicator of 5G networks, and remains a major aspect of 6G networks, too. Indeed, recent studies argue that 5G has not achieved the promised 2000x EE increased, actually increasing the EE only by a factor four \cite{David5G}.  

While RISs have the potential of drastically improving the EE of wireless networks, due to their very limited hardware power consumption, most research contributions on radio resource allocation for RIS-aided networks have focused on maximizing the system rate \cite{Ref15,Hu2021,Abrardo21} or on minimizing the power consumption \cite{Wu2018,Zhou2020}, rather than optimizing the bit-per-Joule EE. 
Only few contributions, on the other hand, have started addressing the issue of radio resource allocation for EE maximization in RIS-based networks. In \cite{Ref9}, the EE of the downlink of an RIS-aided multi-user network is optimized, assuming zero-forcing transmission. It is shown that an RIS can be more energy-efficient than an amplify-and-forward relay. In \cite{ZapTWC2021}, a single-user multiple-input multiple-output (MIMO) link is considered, and it is shown that an RIS can provide significant EE gains even using few transmit and receive antennas. In \cite{You2021}, the trade-off between the EE and spectral efficiency is addressed in the downlink of a MIMO multi-user network, by optimizing the transmit beamforming and RIS coefficients. In \cite{Wang2022}, the EE of a non-orthogonal multiple access network is maximized, assuming successive interference cancellation to remove multi-user interference. In \cite{Le2021} the EE of a cell-free RIS-aided network is optimized by means of sequential optimization, with respect to the transmit beamforming and the RIS phase shifts. In \cite{Gao2021}, the EE of an RIS-aided network employing wireless power transfer is addressed, assuming a transmission over orthogonal carriers and applying an outage rate constraint. In \cite{Xu2022}, a max-min EE maximization problem is tackled by Dinkelbach's method in a RIS-aided heterogeneous network with hardware impairments. 

In this paper, we analyze the issue of EE maximization in an RIS-aided multi-user wireless network, by novel optimization methods. Unlike previous works, the optimization is performed assuming that the RIS is capable of global reflection, i.e. the constraint on the power reflected by the RIS is not applied to each reflecting element individually, but rather to the complete surface \cite{MDR22}. Finally, EE optimization is tackled not only with respect to the transmit powers and RIS reflection coefficients, as it is customary in the literature, but also with respect to the linear receive filters. As described in the rest of the work, this complicates the mathematical structure of the function to maximize, especially when the optimal receive filters are embedded into the EE function.  
\vspace{-0.1cm}
\section{System Model}\label{Sec:Sys}
\begin{figure}[!h]
\centering
\includegraphics[width=0.35\textwidth]{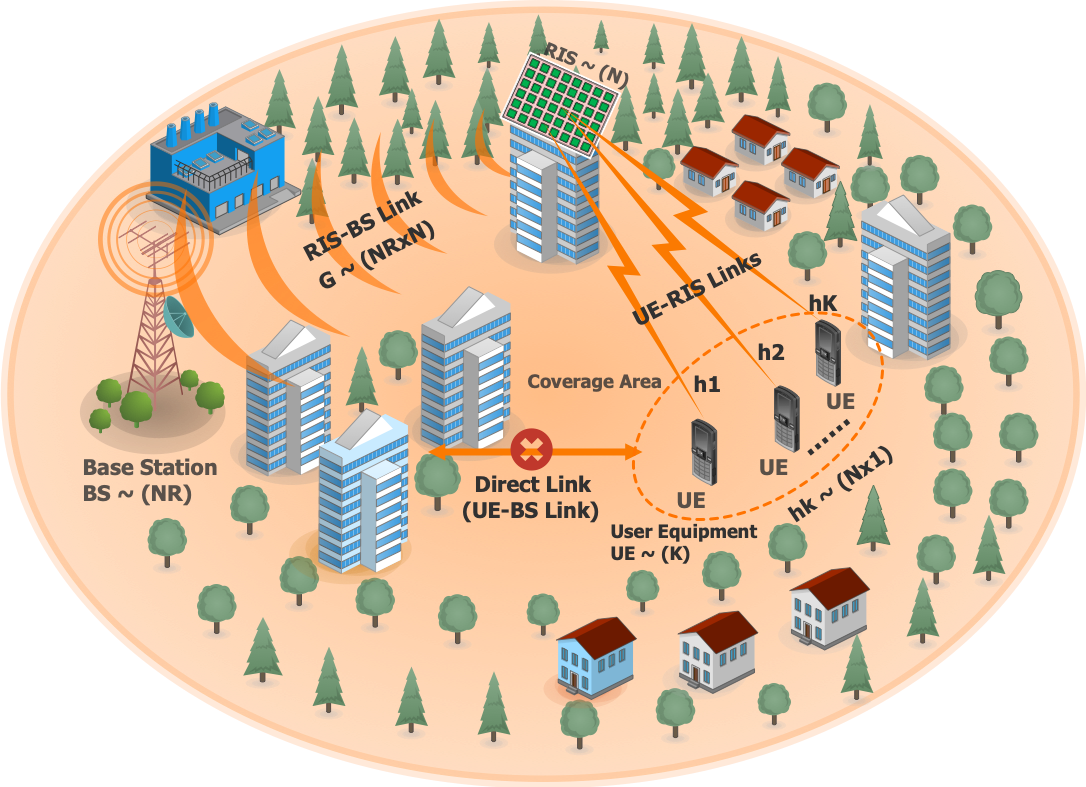}\caption{Considered wireless network.} \label{Fig:scenario}
\end{figure}

\vspace{-0.1cm}
Let us consider the uplink of a multi-user system in which $K$ single-antenna mobile terminals communicate with a base station equipped with $N_{R}$ antennas, through an RIS with $N$ reflecting elements (Fig. \ref{Fig:scenario}). 
Let us denote by $\bh_{k}$ the $N\times 1$ channel from user $k$ to the RIS, by $\bG$ the $N_{R}\times N$ channel from the RIS to the base station, by $\bGamma=\text{diag}(\gamma_{1},\ldots,\gamma_{N})$, the matrix whose diagonal contains the $N$ RIS reflection coefficients, and by $p_{k}$ and $s_{k}$ the transmit power and information symbol of user $k$. 

Related works in the literature consider RISs with local reflection capabilities, i.e. they assume that each RIS element is individually constrained to be less than a threshold, namely $|\gamma_{n}|^{2}\leq P_{R}$ for all $n=1,\ldots,N$, with $P_{R}\leq 1$. Instead, this work considers the more general scenario in which the RIS is characterized by a global reflection constraints, i.e. a constraint on the total power reflected by all of the RIS elements, namely $\sum_{n=1}^{N}|\gamma_{n}|^{2}\leq \sum_{n=1}^{N}P_{R}=NP_{R}$. We should note that in both cases the total power reflected by the RIS is equal to $NP_{R}$. However, the global reflection constraint is more general, as it allows for the modulus of some reflection coefficients to be larger than one, while ensuring that the RIS is still nearly-passive from the global point of view, i.e. the total reflected power is not larger than the total incident power. 

The $k$-th user's signal-to-interference-plus-noise ratio (SINR) after applying the receive filter $\bc_{k}$, is  
\begin{align}\label{Eq:SINR2}
\text{SINR}_{k}&=
\frac{p_{k}|\bc_{k}^{H}\bA_{k}\bgamma|^{2}}{\sigma^{2}\|\bc_{k}\|^{2}+\sum_{m\neq k}p_{m}|\bc_{k}^{H}\bA_{m}\bgamma|^{2}}\;,
\end{align}
where we have defined $\bH_{k}=\text{diag}(\bh_{k}(1),\ldots,\bh_{k}(N))$, $\bgamma=[\gamma_{1},\ldots,\gamma_{N}]$, and $\bA_{k}=\bG\bH_{k}$, for all $k=1,\ldots,K$. Then, the $k$-th user's achievable rate is $R_{k}=B\log_{2}\left(1+\text{SINR}_{k}\right)$, while the system global energy efficiency (GEE) is written as 
\begin{align}
\text{GEE}&=B\frac{\sum_{k=1}^{K}\log_{2}\left(1+\text{SINR}_{k}\right)}{P_{c}+\sum_{k=1}^{K}\mu_{k}p_{k}}\;,
\end{align}
with $B$ the communication bandwidth, $P_{c}=P_{0}+P_{RIS}$ the total hardware power consumed in the system, with $P_{0}$ the hardware power consumed by the base station and mobile terminals, and $P_{RIS}$ the hardware power consumed by the RIS, i.e., $P_{RIS}=NP_{c,n}+P_{0,RIS}$, with $P_{c,n}$ the static power consumed by each RIS element, and $P_{0,RIS}$ accounting for all other sources of static power consumption at the RIS. 

The problem of interest in this work is 
\begin{subequations}\label{Prob:PassiveRIS}
\begin{align}
&\ds\max_{\bgamma,\bp,\bC}\; \text{GEE}(\bgamma,\bp,\bC)\label{Prob:aPassiveRIS}\\
&\;\text{s.t.}\;\|\bgamma\|^{2}\leq NP_{R}\label{Prob:bPassiveRIS}\;,0\leq p_{k}\leq P_{max,k}\;\forall k=1,\ldots,K
\end{align}
\end{subequations}
wherein we have defined $\bp=[p_{1},\ldots,p_{K}]$, $\bC=[\bc_{1},\ldots,\bc_{K}]$, and $P_{max,k}$ is the $k$-th user's maximum transmit power. In order to tackle the non-convex Problem \eqref{Prob:PassiveRIS}, two algorithms are presented in Sections \ref{Sec:Design1} and \ref{Sec:Design2}. 

\vspace{-0.1cm}
\section{First proposed approach}\label{Sec:Design1}
\vspace{-0.1cm}
The first optimization method is based on the alternating optimization algorithm applied to the 
variables $\bgamma$, $\bp$, and $\bC$. 

\textbf{Optimization of $\bgamma$.} Since the vector $\bgamma$ affects only the numerator of the GEE, the optimization of $\gamma$ is stated as 
\begin{align}\label{Prob:MaxGEEMF_gamma}
&\max_{\bgamma}\,\textstyle\sum_{k=1}^{K}R_{k}\;,\text{s.t.}\;\|\bgamma\|^{2}\leq NP_{R}\;.
\end{align}
Problem \eqref{Prob:MaxGEEMF_gamma} is challenging because the objective is not concave in $\bgamma$. In order to provide a practical, but at the same time theoretically grounded method for tackling \eqref{Prob:MaxGEEMF_gamma}, we resort to the sequential approximation method \cite{SeqCvxProg78}. To this end, we need to find a concave lower-bound of \eqref{Prob:MaxGEEMF_gamma}. To this end, we begin by applying the lower-bound $\log_{2}\left(1+\frac{x}{y}\right)\geq \log_{2}\left(1+\frac{\bar{x}}{\bar{y}}\right)+\frac{\bar{x}}{\bar{y}}\left(\frac{2\sqrt{x}}{\sqrt{\bar{x}}}-\frac{x+y}{\bar{x}+\bar{y}}-1\right)$, which holds for any $x$, $y$, $\bar{x}$ and $\bar{y}$, and holds with equality whenever $x=\bar{x}$ and $y=\bar{y}$. Elaborating yields
\begin{align}
&R_{k}\!=\!\log_{2}\!\left(\!1\!+\!\frac{p_{k}|\bc_{k}^{H}\bA_{k}\bgamma|^{2}}{\sigma^{2}\|\bc_{k}\|^{2}+\sum_{m\neq k}p_{m}|\bc_{k}^{H}\bA_{m}\bgamma|^{2}}\right)\geq\label{Eq:BarR}\\
&\bar{A}_{k}\!+\!\bar{B}_{k}\!\left(\!\bar{D}_{k}|\bc_{k}^{H}\bA_{k}\bgamma|\!-\!\bar{E}_{k}\!\!\sum_{m=1}^{K}p_{m}|\bc_{k}^{H}\bA_{m}\bgamma|^{2}\!-\!\bar{F}_{k}\!\right)\!=\!\bar{R}_{k}\notag
\end{align}
with $\bar{\bgamma}$ any feasible vector of RIS coefficients, and $\bar{A}_{k}=\log_{2}\left(1+p_{k}|\bc_{k}^{H}\bA_{k}\bar{\bgamma}|^{2}/I_{k}\right)$, $B_{k}=p_{k}|\bc_{k}^{H}\bA_{k}\bar{\bgamma}|^{2}/I_{k}$, $D_{k}=2/|\bc_{k}^{H}\bA_{k}\bar{\bgamma}|$, $E_{k}=1/I_{k}$, $F_{k}=E_{k}\sigma^{2}\|\bc_{k}\|^{2}+1$, $I_{k}=\sigma^{2}\|\bc_{k}\|^{2}+\sum_{m\neq k}p_{m}|\bc_{k}^{H}\bA_{m}\bar{\bgamma}|^{2}$. The function in \eqref{Eq:BarR} is not concave in $\bgamma$ due to the term $|\bc_{k}^{H}\bA_{k}\bgamma|$, which is convex\footnote{Recall that the composition of a convex function with a convex and increasing function is convex.} in $\bgamma$. However, being convex, $|\bc_{k}^{H}\bA_{k}\bgamma|$ can be lower-bounded by its first-order Taylor expansion around any point. Thus, considering the Taylor expansion of $|\bc_{k}^{H}\bA_{k}\bgamma|$ around $\bar{\bgamma}$, we obtain the concave lower-bound  
\begin{align}
R_{k}&\!\geq\!\bar{R}_{k}\!\geq\!\bar{B}_{k}\!\Bigg(\!\bar{D}_{k}\!\left(|\bc_{k}^{H}\bA_{k}\bar{\bgamma}|+\Re\!\left\{\!\frac{\bA_{k}^{H}\bc_{k}\bc_{k}^{H}\bA_{k}\bar{\gamma}}{|\bc_{k}^{H}\bA_{k}\bar{\bgamma}|}(\bgamma\!-\!\bar{\bgamma})\!\right\}\right)\notag\\
\!&-\!\bar{E}_{k}\!\sum_{m=1}^{K}\!p_{m}|\bc_{k}^{H}\bA_{m}\bgamma|^{2}-\bar{F}_{k}\Bigg)+\bar{A}_{k}\!=\!\widetilde{R}_{k}(\bgamma)
\end{align}
Thus, in each iteration of the sequential method, the following convex problem is to be solved 
\begin{subequations}
\begin{align}
&\max_{\bgamma}\textstyle\sum_{k=1}^{K}\widetilde{R}_{k}(\bgamma)\;,\;\text{s.t.}\;\|\bgamma\|^{2}\leq NP_{R}
\end{align}
\end{subequations}

\textbf{Optimization of $\bp$.} Defining $a_{k,m}=|\bc_{k}^{H}\bA_{m}\bgamma|^{2}$, for all $m$ and $k$, and $d_{k}=\sigma^{2}\|\bc_{k}\|^{2}$, the power optimization problem can be stated as
\begin{subequations}\label{Prob:MaxGEEMF_power}
\begin{align}
&\ds\max_{\bp}\,\frac{\sum_{k=1}^{K}\ds\log_{2}\left(1+\frac{p_{k}a_{k,k}}{d_{k}+\sum_{m\neq k}p_{m}a_{k,m}}\right)}{\sum_{k=1}^{K}\mu_{k}p_{k}+P_{c}^{(p)}}\label{Prob:aMaxGEEMF_power}\\
&\;\text{s.t.}\;0\leq p_{k}\leq P_{max,k}\;,\forall\; k=1,\ldots,K
\end{align}
\end{subequations}
Since the numerator of \eqref{Prob:aMaxGEEMF_power} is not a concave function of $\bp$, the objective in \eqref{Prob:aMaxGEEMF_power} is not a pseudo-concave function and thus it is computationally unfeasible to solve \eqref{Prob:MaxGEEMF_power} by fractional programming  \cite{ZapNow15}. Then, we resort to the sequential fractional programming method \cite{ZapNow15}, in order to derive a pseudo-concave lower-bound of \eqref{Prob:aMaxGEEMF_power}, which can be maximized by fractional programming. To this end, we express \eqref{Prob:aMaxGEEMF_power} as  
\begin{align}
\text{GEE}(\bp)&=\underbrace{\frac{\sum_{k=1}^{K}\log_{2}\left(\!d_{k}+\sum_{k=1}^{K}p_{k}a_{k,k}\right)}{\sum_{k=1}^{K}\mu_{k}p_{k}+P_{c}^{(p)}}}_{\ds g_{1}(\bp)}\\
&-\underbrace{\frac{\sum_{k=1}^{K}\log_{2}\left(d_{k}+\sum_{m\neq k}p_{m}a_{k,m}\right)}{\sum_{k=1}^{K}\mu_{k}p_{k}+P_{c}^{(p)}}}_{\ds g_{2}(\bp)}.\notag
\end{align}
Then, a pseudo-concave lower-bound of $\text{GEE}(\bp)$, which we denote by $\widetilde{\text{GEE}}(\bp)$, can be obtained by replacing the numerator of $g_{2}(\bp)$ with its first-order Taylor expansion around any feasible point $\bar{p}$ \cite{ZapNow15}.
Thus, in each iteration of the sequential method, the problem to be solved is the maximization of the pseudo-concave function $\widetilde{\text{GEE}}(\bp)$, subject to the power constraint $p_{k}\in[0,P_{max,k}]$, for all $k=1,\ldots,K$, which can be globally and efficiently solved by fractional programming. 

\textbf{Optimization of $\bC$.} The optimization of the receive filters affects again only the numerator of the GEE. Moreover, it can be decoupled over the users, thus reducing the problem to the maximization of the individual rates of the users. The solution to this problem is well-known to be the linear minimum mean squared error (MMSE) receiver, which for the case at hand, is expressed as $\bc_{k}=\sqrt{p}_{k}\bM_{k}^{-1}\bA_{k}\bgamma$, with $\bM_{k}=\sum_{m\neq k}p_{m}\bA_{m}\bgamma\bgamma^{H}\bA_{m}^{H}+\sigma^{2}\bI_{N_{R}}$. 
\vspace{-0.1cm}
\section{Second proposed approach}\label{Sec:Design2}
\vspace{-0.1cm}
While the first proposed approach considers the alternating optimization of three variables, namely, the transmit powers $\bp$, the receive filters $\bc_{1},\ldots,\bc_{K}$, and the RIS reflection coefficients $\bgamma$, in this section we develop an algorithm that embeds the optimal expression of the linear MMSE filters into the GEE, and optimizes the resulting expression with respect to $\bp$ and $\bgamma$. While this is expected to yield better performance, an intuition that will be confirmed by numerical results, it also leads to a more challenging optimization problem, due to the more involved expression of the EE function. 

To elaborate, plugging the expression of the linear MMSE receive filters, i.e. $\bc_{k}=\sqrt{p}_{k}\bM_{k}^{-1}\bA_{k}\bgamma$, into \eqref{Eq:SINR2} leads to 
\begin{align}
&\text{SR}_{\text{MMSE}}=\textstyle\sum_{k=1}^{K}\log_{2}\!\left(1\!+\!p_{k}\bgamma^{H}\bA_{k}^{H}\bM_{k}^{-1}\bA_{k}\bgamma\!\right)\label{Eq:SR_MMSE}=\\
&\textstyle\sum_{k=1}^{K}\!\log_{2}\!\left|\!\sigma^{2}\bI_{N_{R}}\!+\!\sum_{m=1}^{K}p_{m}\bA_{m}\bgamma\bgamma^{H}\bA_{m}^{H}\!\right|\!-\!\sum_{k=1}^{K}\!\log_{2}\!\left|\bM_{k}\right|\notag
\end{align}
Thus, the GEE maximization problem becomes 
\begin{subequations}\label{Prob:GEEMax}
\begin{align}
&\ds\max_{\bgamma,\bp}\;\frac{\text{SR}_{\text{MMSE}}(\bgamma,\bp)}{\sum_{k=1}^{K}\mu_{k}p_{k}+P_{c}^{(p)}}\label{Prob:aGEEMax}\\
&\;\text{s.t.}\;\|\bgamma\|^{2}\!\leq \!NP_{R},\;p_{k}\in[0,P_{max,k}],\forall k=1,\ldots,K\label{Prob:cGEEMax}
\end{align}
\end{subequations}
Problem \eqref{Prob:GEEMax} is tackled by alternating optimization of $\bp$ and $\bgamma$, as discussed next. 

\textbf{Optimization of $\bgamma$.} Since $\bgamma$ affects only the numerator of \eqref{Prob:aGEEMax}, when $\bp$ is fixed the problem reduces to the maximization of the system sum-rate, which is given by $\text{SR}_{\text{MMSE}}(\bgamma)$ in \eqref{Eq:SR_MMSE}, subject to $\|\bgamma\|^{2}\leq NP_{R}$. The objective is not concave, but we can proceed by defining $\bX=\bgamma\bgamma^{H}$, $\bB_{k}=p_{k}\bA_{k}\bX\bA_{k}^{H}$, and $\bB=\sum_{k=1}^{K}\bB_{k}$, which yields 
\vspace{-0.3cm}
\begin{align}\label{Eq:SR_X}
\text{SR}_{\text{MMSE}}(\bX)&\!=\!\!\underbrace{\sum_{k=1}^{K}\log_{2}\!\left|\!\sigma^{2}\bI_{N_{R}}\!+\!\bB\!\right|}_{\ds G_{1}(\bX)}\!-\!\underbrace{\sum_{k=1}^{K}\log_{2}\!\left|\sigma^{2}\bI_{N_{R}}\!+\!\!\sum_{m\neq k}\!\bB_{m}\right|}_{\ds G_{2}(\bX)},\notag
\end{align}
\noindent which can be lower-bounded by a concave function by expanding $G_{2}(\bX)$ around any point $\bar{\bX}$, i.e. 
$\text{SR}_{\text{MMSE}}(\bX)\geq G_{1}(\bX)-G_{2}(\bar{\bX})\!-\!\!\Re\left\{\tr\left(\!\nabla G_{2}(\bar{\bX})^{H}(\bX\!-\!\bar{\bX})\!\right)\!\right\}\!=\!\widetilde{\text{SR}}_{\text{MMSE}}$. Then, relaxing the rank-one constraint on $\bX$, the resulting problem in each iteration of the sequential method is
\vspace{-0.2cm}
\begin{align}
&\ds\max_{\bX}\,\widetilde{\text{SR}}_{\text{MMSE}}(\bX)\;,\text{s.t.}\;\tr(\bX)\leq NP_{R}\;,\bX\succeq \bzero\;.
\end{align}
Upon convergence of the sequential procedure, if the convergence point $\bX^{*}$ has unit-rank, it is  feasible for the original problem. Otherwise, a feasible solution can be obtained by randomization or rank reduction techniques \cite{LuoSDR}. 

\textbf{Optimization of $\bp$.} When $\bgamma$ is fixed, the problem 
can be tackled by resorting to the sequential fractional programming framework. Indeed, the sum-rate in \eqref{Eq:SR_MMSE} is the difference of two concave functions of $\bp$, and thus the GEE in \eqref{Prob:aGEEMax} can be lower-bounded by linearizing the negative term in \eqref{Eq:SR_MMSE}, i.e. the function $F(\bp)\!=\!\ds\sum_{k=1}^{K}\log_{2}\!\left|\!\sigma^{2}\bI_{N_{R}}\!+\!\ds\sum_{m\neq k}p_{m}\bA_{m}\bgamma\bgamma^{H}\bA_{m}^{H}\right|$, around any feasible point $\bar{\bp}$. This leads to
\begin{align}
&\text{GEE}_{\text{MMSE}}(\bp)\geq\frac{\sum_{k=1}^{K}\log_{2}\left|\sigma^{2}\bI_{N_{R}}+\sum_{m=1}^{K}p_{m}\bA_{m}\bgamma\bgamma^{H}\bA_{m}^{H}\right|}{\sum_{k=1}^{K}\mu_{k}p_{k}+P_{c}^{(p)}}\notag\\
&-\frac{F(\bar{\bp})}{\sum_{k=1}^{K}\mu_{k}\bar{p}_{k}+P_{c}^{(p)}}-\frac{\left(\nabla_{\bp} F(\bar{\bp})\right)^{T}\left(\bp-\bar{\bp}\right)}{\sum_{k=1}^{K}\mu_{k}p_{k}+P_{c}^{(p)}}=\widetilde{\text{GEE}}_{\text{MMSE}}\notag
\end{align}
Thus, in each iteration of the sequential fractional programming algorithm, the problem to be solved is stated as  
\begin{align}\label{Prob:GEEMax_SFP}
&\!\!\!\ds\max_{\bp}\widetilde{\text{GEE}}_{\text{MMSE}}(\bp),\text{s.t.}\,p_{k}\!\in\![0,P_{max,k}],\forall k\!=\!1,\ldots,K,
\end{align}
which has linear constraints and whose objective is the ratio between a concave and an affine function \cite{ZapNow15}. Thus, \eqref{Prob:GEEMax_SFP} can be solved by fractional programming. 
\vspace{-0.2cm}
\section{Numerical Results}\label{Sec:Numerics}
\vspace{-0.2cm}
We consider an instance of the network described in Section \ref{Sec:Sys}, with $K=4$, $N_{R}=4$, $N=100$, $B=20\,\textrm{MHz}$, $P_{0}=40\,\textrm{dBm}$, $P_{0,RIS}=20\,\textrm{dBm}$, $P_{c,n}=0\,\textrm{dBm}$. The noise power spectral density is $-174\,\textrm{dBm/Hz}$, and the noise figure is $10\,\textrm{dB}$. The users are randomly placed in an area with radius $100\,\textrm{m}$ around the RIS, and the base station is placed $50\,\textrm{m}$ away from the RIS. The users are at a random height in $[0,5]\,\textrm{m}$, while the RIS and base station are at heights of $15\,\textrm{m}$ and $10\,\textrm{m}$, respectively. The path-loss exponent is $\eta=4$, and Rice fading is considered for all channels, with factors $K_{t}=4$ for the channel from the RIS to the base station and $K_{r}=2$ for the channels from the mobile users to the RIS. 

\begin{figure}[!h]
\centering
\includegraphics[width=0.43\textwidth]{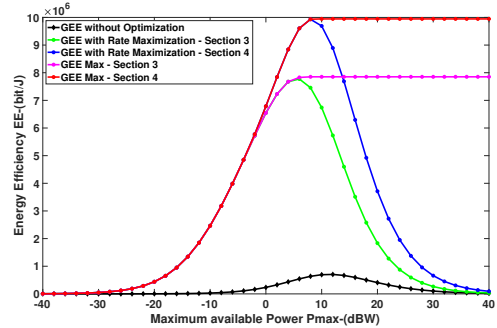}\caption{EE versus $P_{max}$. $K=4$, $N_{R}=4$, $N=100$.} \label{fig:EEvsP}
\end{figure}

\begin{figure}[!h]
\centering
\includegraphics[width=0.43\textwidth]{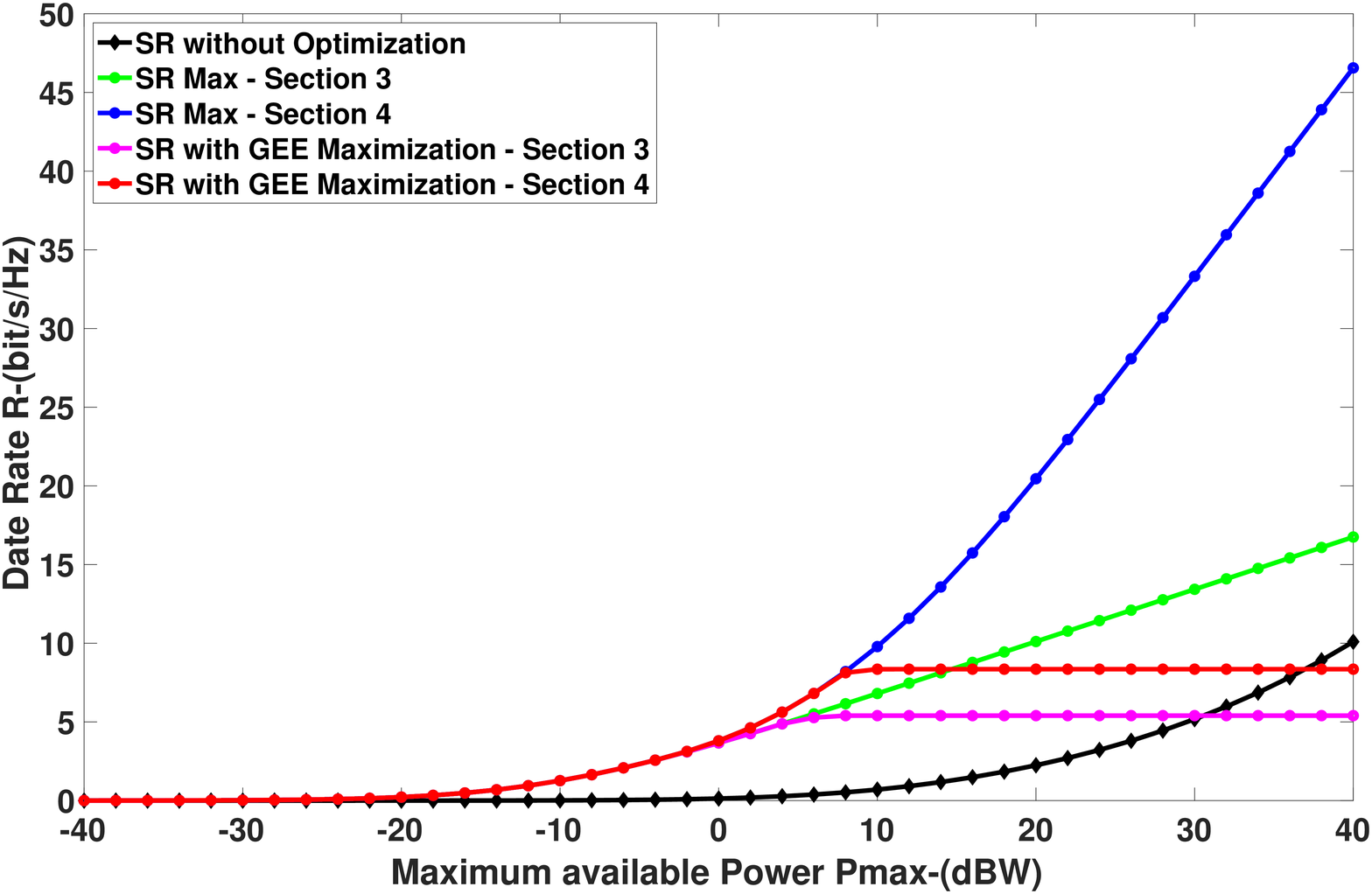}\caption{Rate versus $P_{max}$. $K=4$, $N_{R}=4$, $N=100$.} \label{fig:RvsP}
\end{figure}

\begin{figure}[!h]
\centering
\includegraphics[width=0.43\textwidth]{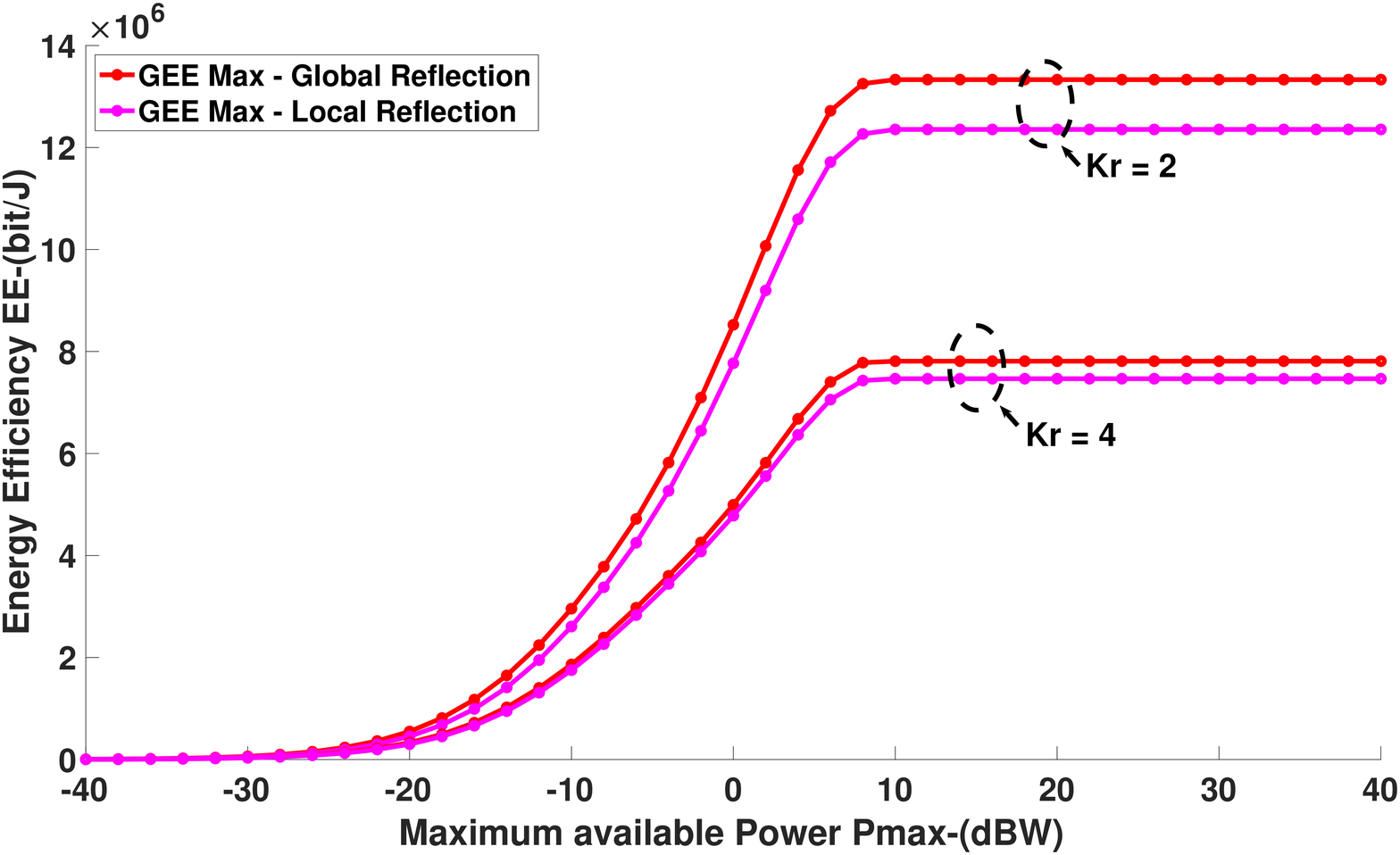}\caption{GEE versus $P_{max}$. $K=4$, $N_{R}=4$, $N=100$.} \label{fig:EEvsK}
\end{figure}

Figure \ref{fig:EEvsP} shows the GEE achieved by: (a) maximizing the GEE by the method from Section \ref{Sec:Design1}; (b) maximizing the GEE by the method from Section \ref{Sec:Design2}; (c) the resource allocation obtained by adapting the method from Section\footnote{Both the methods from Sections \ref{Sec:Design1} and \ref{Sec:Design2} can be be specialized to perform sum-rate maximization by simply setting $\mu_{k}=0$, for all $k=1,\ldots,K$.} \ref{Sec:Design1} for rate maximization; (d) the resource allocation obtained by adapting the method from Section \ref{Sec:Design2} for rate maximization; (e) uniform power allocation and random RIS phase shifts. As anticipated, the method from Section \ref{Sec:Design2} significantly outperforms the approach from Section \ref{Sec:Design1}, thanks to the exploitation of the mathematical structure of the optimal receive filter, rather than simply updating the receive filter within the alternating maximization algorithm. Moreover, a large gain is obtained compared to Case (e), in which radio resources are not optimized. Figure \ref{fig:RvsP} shows the rate achieved by: (a) maximizing the rate by the method from Section \ref{Sec:Design1}; (b) maximizing the rate by the method from Section \ref{Sec:Design2}; (c) the resource allocation obtained by adapting the method from Section \ref{Sec:Design1} for GEE maximization; (d) the resource allocation obtained by adapting the method from Section \ref{Sec:Design2} for GEE maximization; Similar considerations as for Fig. \ref{fig:EEvsP} can be made. Finally, Fig. \ref{fig:EEvsK} compares the maximum GEE achieved by the method from Section \ref{Sec:Design2} to the special case in which the same GEE maximization method from Section \ref{Sec:Design2} is specialized to the benchmark scheme of a RIS with local reflection constraints, i.e. replacing \eqref{Prob:bPassiveRIS} by $|\gamma_{n}|\leq P_{R}$, for all $n=1,\ldots,N$. The comparison is made for $K_{t}=K_{r}=2;4$. As expected, the results show that a global reflection constraint provides a visible gain, since it widens the set of feasible RIS configurations. Also, the gain is more significant for lower values of the Rician factor. Indeed, in this case the channel realizations tend to be more spread out and an RIS with global reflections can better adapt to this scenario.

\vspace{-0.3cm}
\section{Conclusions}\label{Sec:Conclusion}
\vspace{-0.1cm}
This work addressed the EE maximization problem in a multi-user network aided by an RIS endowed with global reflection capabilities. The results indicate that the proposed radio resource optimization algorithms provide large EE gains compared to heuristic random resource allocations. Moreover, a careful optimization of the receive filters and exploitation of the RIS global reflection capabilities can provide significant performance improvements. 

\bibliographystyle{IEEEbib}
\bibliography{FracProg,references}

\end{document}

%% file: abbreviations.tex
\newcommand{\eeq}{\end{equation}}

\newcommand{\bp}{\mbox{\boldmath $p$}}

\newcommand{\bzero}{\mbox{\boldmath $0$}}
\newcommand{\bM}{\mbox{\boldmath $M$}}

\newcommand{\bA}{\mbox{\boldmath $A$}}

\newcommand{\bC}{\mbox{\boldmath $C$}}
\newcommand{\bX}{\mbox{\boldmath $X$}}
\newcommand{\bh}{\mbox{\boldmath $h$}}
\newcommand{\bH}{\mbox{\boldmath $H$}}

\newcommand{\bc}{\mbox{\boldmath $c$}}

\newcommand{\bgamma}{\mbox{\boldmath $\gamma$}}

\newcommand{\bGamma}{\mbox{\boldmath $\Gamma$}}
\newcommand{\bB}{\mbox{\boldmath $B$}}

\newcommand{\bG}{\mbox{\boldmath $G$}}

\newcommand{\bI}{\mbox{\boldmath $I$}}

\newcommand{\ds}{\displaystyle}

\newcommand{\beq}{\begin{equation}}

\newcommand{\tr}      {{\mathrm{tr}}}    



%% file: acronyms.tex
\makeglossaries

\newacronym{mac}{MAC}{multiple-access channel}
\newacronym{bc}{BC}{broadcast channel}
\newacronym{mimo}{MIMO}{multiple-input multiple-output}
\newacronym{siso}{SISO}{single-input single-output}
\newacronym{sc}{SC}{single-carrier}
\newacronym{mc}{MC}{multi-carrier}
\newacronym{ofdma}{OFDMA}{orthogonal frequency division multiple access}
\newacronym{af}{AF}{amplify-and-forward}
\newacronym{df}{DF}{decode-and-forward}
\newacronym{cf}{CF}{compress-and-forward}
\newacronym{mwrc}{MWRC}{multi-way relay channel}
\newacronym{pde}{PDE}{partial data exchange}
\newacronym{fde}{FDE}{full data exchange}
\newacronym{iid}{i.i.d.\@}{independent and identically distributed}
\newacronym{awgn}{AWGN}{additive white Gaussian noise}
\newacronym{awg}{AWG}{additive white Gaussian}
\newacronym{sic}{SIC}{successive interference cancellation}
\newacronym{snr}{SNR}{signal-to-noise ratio}
\newacronym{sinr}{SINR}{signal-to-interference-plus-noise ratio}
\newacronym{ber}{BER}{bit error rate}
\newacronym{zf}{ZF}{zero-forcing}
\newacronym{mmse}{MMSE}{minimum mean square error}
\newacronym{sud}{SUD}{single user decoding}
\newacronym{dof}{DoF}{degrees of freedom}
\newacronym{gdof}{GDoF}{generalized degrees of freedom}
\newacronym{nnc}{NNC}{noisy network coding}
\newacronym{dmn}{DMN}{discrete memoryless network}
\newacronym{csi}{CSI}{channel state information}
\newacronym{ee}{EE}{energy efficiency}
\newacronym{ian}{IAN}{treating interference as noise}
\newacronym{snd}{SND}{simultaneous non-unique decoding}
\newacronym{brd}{BRD}{best response dynamics}
\newacronym{br}{BR}{best response}
\newacronym{ne}{NE}{Nash equilibrium}
\newacronym{lhs}{LHS}{left-hand side}
\newacronym{rhs}{RHS}{right-hand side}
\newacronym{gee}{GEE}{global energy efficiency}
\newacronym{wsee}{WSEE}{weighted sum energy efficiency}
\newacronym{wpee}{WPEE}{weighted product energy efficiency}
\newacronym{wmee}{WMEE}{weighted minimum energy efficiency}
\newacronym{kkt}{KKT}{Karush-Kuhn-Tucker}
\newacronym{pc}{PC}{pseudo-concave}
\newacronym{qc}{QC}{quasi-concave}
\newacronym{ql}{QL}{quasi-linear}
\newacronym{pl}{PL}{pseudo-linear}
\newacronym{spc}{SPC}{strictly pseudo-concave}
\newacronym{sqc}{SQC}{strictly quasi-concave}
\newacronym{lfp}{LFP}{linear fractional problem}
\newacronym{clfp}{CLFP}{concave-linear fractional problem}
\newacronym{ccfp}{CCFP}{concave-convex fractional problem}
\newacronym{mmfp}{MMFP}{max-min fractional problem}
\newacronym{sorp}{SoRP}{sum-of-ratios problem}
\newacronym{porp}{PoRP}{product-of-ratios problem}
\newacronym{srp}{SRP}{single-ratio problem}
\newacronym{brb}{BRB}{branch-reduce-and-bound}
\newacronym{qos}{QoS}{quality-of-service}
\newacronym{comp}{CoMP}{cooperative multi-point}
\newacronym{ue}{UE}{user equipment}
\newacronym{bs}{BS}{base station}
\newacronym{mrc}{MRC}{maximum ratio combining}
\newacronym{d2d}{D2D}{device-to-device}
\newacronym{lmmse}{LMMSE}{linear minimum mean square error}
\newacronym{ris}{RIS}{reconfigurable intelligent surface}
\newacronym{svd}{SVD}{singular values decomposition}